\documentclass[aps,preprint,amsmath,amssymb,superscriptaddress,showpacs]{revtex4}
\usepackage{graphicx}
\begin{document}
\title{Self-consistent Overhauser model for the pair distribution function
of an electron gas at finite temperature}
\author{R. Asgari}
\affiliation{NEST-INFM and Classe di Scienze, Scuola Normale Superiore, I-56126 Pisa, Italy}
\affiliation{Institute for Studies in Theoretical Physics and Mathematics, Tehran 19395-5531, Iran}
\author{M. Cardenas}
\affiliation{NEST-INFM and Classe di Scienze, Scuola Normale Superiore, I-56126 Pisa, Italy}
\affiliation{Instituto de F\'isica, Universidad Cat\'olica de Valpara\'iso, 2374638 Valpara\'iso, Chile}
\author{M. Polini}
\affiliation{NEST-INFM and Classe di Scienze, Scuola Normale Superiore, I-56126 Pisa, Italy}
\author{B. Davoudi\footnote{Present address: D\'epartement de Physique and Centre de Recherche en Physique du Solide, 
Universit\'e de Sherbrooke, Sherbrooke, Qu\'ebec, Canada J1K 2R1}}
\affiliation{NEST-INFM and Classe di Scienze, Scuola Normale Superiore, I-56126 Pisa, Italy}
\affiliation{Institute for Studies in Theoretical Physics and Mathematics, Tehran 19395-5531, Iran}
\author{M.P. Tosi}
\affiliation{NEST-INFM and Classe di Scienze, Scuola Normale Superiore, I-56126 Pisa, Italy}

\begin{abstract}
We present calculations of the spin-averaged pair distribution function $g(r)$ in a 
homogeneous gas of electrons moving in dimensionality $D=3$ or $D=2$ at finite temperature. 
The model involves the solution of a two-electron scattering problem via an effective potential 
which embodies many-body effects through a self-consistent Hartree approximation, leading to 
two-body wave functions to be averaged over a temperature-dependent distribution of relative 
momentum for electron pairs. We report illustrative numerical results for $g(r)$ in an 
intermediate-coupling regime and interpret them in terms of changes of short-range order with 
increasing temperature.
\end{abstract}
\maketitle

\section{Introduction}
Many of the electron-electron interaction effects in condensed-matter and astrophysical 
plasmas can be understood by reference to the homogeneous electron-gas model. A central role 
is played by the electron-pair distribution function $g(r)$, which describes how short-range 
exchange and correlations enter to determine the probability of finding two electrons at a relative 
distance $r$. Most of the theoretical and computational studies of this model have been concerned 
either with the limit of complete quantum degeneracy at zero temperature~\cite{1} or with the classical 
limit at very high temperature~\cite{2}. In a number of physical situations, however, the state of the 
electrons is in neither of these limiting regimes. Some studies of the equation of state of the 
three-dimensional ($3D$) electron gas in the temperature-density plane were reported most notably 
by Ichimaru and coworkers~\cite{3} and by Macke and coworkers~\cite{4}. In fact, the excess free energy 
of the electron gas at any temperature T can be calculated from the temperature and density 
dependence of $g(r)$ by integration over the coupling strength parameter. 

Accurate and theoretically well founded methods have been developed in recent years for 
the calculation of $g(r)$ in both the $3D$ and the $2D$ electron gas at $T=0$. Here we are specifically 
concerned with the approach originally proposed by Overhauser~\cite{5} for the evaluation of the 
value $g(0)$ of the pair function at contact. Writing $g(r)$ in the paramagnetic state as the average of 
the distributions for parallel- and antiparallel-spin electron pairs, $g(r)=[g_{\uparrow\uparrow}(r)+g_{\uparrow\downarrow}(r)]/2$, one 
has that $g_{\uparrow\uparrow}(0)$ vanishes on account of the Pauli principle and $g_{\uparrow\downarrow}(0)$ 
is solely determined by two-body scattering events. In Overhauser's model $g_{\uparrow\downarrow}(0)$ 
was obtained from the solution of a Schr\"odinger equation for the relative motion of two electrons with antiparallel spins in an effective interaction potential which also accounts for many-body effects from the electron gas. 
This approach was later extended to evaluate $g(r)$ at finite $r$ as a sum of contributions from two-
body wave functions (``geminals"), the geminal weights being determined by the distribution of 
relative momentum for a pair of electrons in the electron gas~\cite{6,7,8,9}.

In the present work we extend to finite temperature the approach of Davoudi {\it et al}.~\cite{7}, 
which was based on a self-consistent Hartree model for the effective scattering potential. Section 
2 summarizes the method used for the evaluation of the pair wave functions from the solution of 
an effective two-body Schr\"odinger equation in dimensionality $D=3$ and $D=2$, while Section 3 
presents the calculation of the temperature-dependent geminal weights needed to construct the 
pair distribution function. Some illustrative numerical results for $g(r)$ are presented and 
discussed in Section 4, and finally Section 5 offers a brief summary and some perspectives for 
future work.

\section{Two-body wave functions}
The thermodynamic variables that characterize the states of the gas in dimensionality $D$ are 
the electron density $n$ and the temperature $T$. It is convenient to introduce the Wigner-Seitz 
density parameter $r_s$ by the usual relation $r_s a_B=[D/(n\Omega_D)]^{1/D}$, where $a_B$ is the Bohr radius and 
$\Omega_D$ is the solid angle in $D$ dimensions ($\Omega_3=4\pi$ and $\Omega_2=2\pi$). We shall measure the thermal 
energy in ${\rm eV}$ units by introducing a reduced temperature $\Theta=k_BT/{\rm eV}$.

As in the gas at zero temperature~\cite{7}, we aim to solve the two-electron scattering problem 
with some effective interaction potential $V(r)$ in order to determine the wave functions $\Psi^{\uparrow\uparrow}_{{\bf k},{\bf K}}({\bf r},{\bf R})$ and $\Psi^{\uparrow\downarrow}_{{\bf k},{\bf K}}({\bf r},{\bf R})$ for the parallel- and antiparallel-spin states of an electron pair. Here, 
${\bf r}={\bf r}_1-{\bf r}_2$ and ${\bf R}=({\bf r}_1+{\bf r}_2)/2$ 
are the relative coordinate and the center-of-mass coordinate of 
the pair, the conjugate momenta being ${\bf k}=({\bf k}_1-{\bf k}_2)/2$ and ${\bf K}$. 
The spin-resolved pair distribution functions $g_{\sigma\sigma'}(r)$ can then be obtained as
\begin{equation}\label{eq1}
g_{\sigma \sigma'}(r)=\langle\langle 
|\Psi_{{\bf k}, {\bf K}}^{\sigma \sigma'}({\bf r}, {\bf R})|^2\rangle_{\theta} \rangle_{p(k)}
\end{equation}
where $\langle\langle...\rangle_{\theta} \rangle_{p(k)}$ represents averages over the scattering angle $\theta$ 
and over the probability $p(k)$ of finding two electrons with relative momentum $k$. Equation~(\ref{eq1}) 
can also be written as a double sum of the radial components of the pair wave functions over angular momenta and relative momentum, the weight of each term in the latter sum being determined by the function $p(k)$~\cite{9}. 

Carrying out the angular-momentum expansions of the pair wave functions and 
performing the angular averages leads to the results
\begin{equation}\label{eq2}
g_{\uparrow \uparrow}(r)=\frac{2}{r^{D-1}}\,\sum_{\ell=1\,\,({\mathrm odd})}^\infty\,{\mathcal B}^{(D)}_{\ell}\,\langle |\Phi^{(D)}_{\ell, {\bf k}}(r)|^2 \rangle_{p(k)}\, 
\end{equation}
and
\begin{equation}\label{eq3}
g_{\uparrow \downarrow}(r)=\frac{1}{r^{D-1}}\left[\langle|\Phi^{(D)}_{0, {\bf k}}(r)|^2\rangle_{p(k)}+\sum_{\ell=1}^\infty\,{\mathcal B}^{(D)}_{\ell}\,\langle |\Phi^{(D)}_{\ell, {\bf k}}(r)|^2 \rangle_{p(k)} \right]\,,
\end{equation}
where ${\mathcal B}^{(3)}_{\ell}=2\ell+1$ and ${\mathcal B}^{(2)}_{\ell}=2$. 
In Eq.~(\ref{eq2}) the sum over $\ell$ runs over odd integers, because the 
spinor associated with the $\uparrow\uparrow$ state is symmetric and hence the coordinate part of the wave 
function is antisymmetric under exchange. The spinor associated with the $\uparrow\downarrow$ state has instead 
no definite symmetry and hence the sum in Eq.~(\ref{eq3}) runs over all integer values of $\ell$. 
The Schr\"odinger equation obeyed by the radial wave functions $\Phi^{(D)}_{\ell,{\bf k}}(r)$ 
is
\begin{equation}\label{eq4}
\left[-\frac{\hbar^2}{m} \frac{d^2}{d r^2} +\frac{\hbar^2\Lambda_{D}}{m\,r^2}+V(r)\right]\Phi^{(D)}_{\ell, {\bf k}}(r)=\frac{\hbar^2k^2}{m}\,\Phi^{(D)}_{\ell, {\bf k}}(r)
\end{equation}
where $\Lambda_{D}={\bf L}^2_D+(D-1)(D-3)/4$, with ${\bf L}^2_3=\ell(\ell+1)$ $(\ell=0,1,2,...)$ and ${\bf L}^2_2=\ell^2$ $(\ell=0,\pm 1,\pm 2,...$: the negative values of $\ell$ are accounted for by the choice ${\mathcal B}^{(2)}_{\ell}=2$ in Eqs.~(\ref{eq2}) and~(\ref{eq3})). In solving Eq.~(\ref{eq4}) we impose as a boundary condition that $\Phi^{(D)}_{\ell,{\bf k}}(r)$ tends asymptotically to the free solution corresponding to $V(r)=0$, except for a phase shift.

In our calculations the scattering potential $V(r)$ is self-consistently constructed in the 
Hartree approximation. That is,
\begin{equation}\label{eq5}
V(r)=v(r)+n\int d^D{\bf r}'v(|{\bf r}-{\bf r}'|)\,[g(r')-1]
\end{equation}
where $v(r)=e^2/r$. The system of equations~(\ref{eq2})-(\ref{eq5}) is to be solved self-consistently after 
determining the momentum distribution $p(k)$ as a function of temperature, as we do in the next 
Section within a free-electron model.

The theory satisfies a number of exact properties of $g(r)$, as discussed in more detail in 
Ref.~\onlinecite{7}: (i) the form of Eq.~(\ref{eq2}) ensures that the relation $g_{\uparrow\uparrow}(0)=0$ 
is satisfied, since all functions $\Phi^{(D)}_{\ell,{\bf k}}(r)$ 
vanish at the origin for $\ell \neq 0$; (ii) the cusp condition on $g_{\uparrow\downarrow}(r)$ 
at the origin~\cite{10} is satisfied if $V(r)$ tends to the bare Coulomb potential for $r\rightarrow 0$, 
as imposed in Eq.~(\ref{eq5}); 
(iii) the normalization of the wave functions and of the probability function $p(k)$ ensures that the 
charge neutrality condition
\begin{equation}\label{eq6}
n\,\int d^D{\bf r}\,\left[g(r)-1\right]=-1
\end{equation}
is satisfied within a numerical accuracy which is determined by the number of angular-
momentum states that are taken into account; and finally (iv) the Hartree-Fock expression
\begin{equation}\label{eq7}
g^{\rm HF}_{\sigma\sigma'}(r)=1-\delta_{\sigma\sigma'}
\left|\frac{2}{n}\int \frac{d^D {\bf q}}{(2\pi)^D}\,n_{\rm FD}(q)\exp{(i{\bf q}\cdot{\bf r})}\right|^2
\end{equation}
is correctly recovered in the weak-coupling limit by setting $V(r)=0$ and by describing the 
occupation number of a state with momentum ${\bf q}$ by the Fermi-Dirac distribution function,
\begin{equation}\label{eq8}
n_{\rm FD}(q)=\left[\exp{\left(\frac{\varepsilon_{\bf q}-\mu_D}{k_B T}\right)}+1\right]^{-1}\,.
\end{equation}
Here, $\varepsilon_{\bf q}=\hbar^2{\bf q}^2/(2m)$ 
is the single-particle energy and $\mu_D$ is the chemical potential, to be 
determined from the mean density $n$ through the relation
\begin{equation}\label{eq9}
n=\frac{2\Omega_D}{(2\pi)^D}\int_0^{+\infty}q^{D-1}\,n_{\rm FD}(q)\,dq\,.
\end{equation}
\section{Distribution of relative momentum}

The probability distribution $p(k)$ 
of finding two electrons with relative momentum $k$ is 
given by
\begin{equation}\label{eq10}
p(k)=\frac{2^{D+2}\Omega_D\,k^{D-1}}{n^2\,(2 \pi)^D}\,\int \,\frac{d^{D}{\bf q}}{(2\pi)^D}\, n_{\rm FD}(q)
n_{\rm FD}(|{\bf q}+2{\bf k}|)\,.
\end{equation}
The prefactor in Eq.~(\ref{eq10}) is obtained from the normalization condition $\int_0^{+\infty}p(k)dk=1$, 
and the factor $2{\bf k}$ in the argument of the second Fermi-Dirac function takes care of the fact that ${\bf k}$ is the 
relative momentum of the two colliding electrons. 

At $T=0$ the chemical potential is equal to the Fermi energy $\varepsilon_{F}=\hbar^2 k^2_F/(2m)$, 
with $k_F=(\alpha_D r_s a_B)^{-1}$ being the Fermi wave number ($\alpha_3=(9\pi/4)^{-1/3}$ and $\alpha_2=2^{-1/2}$), 
and simple analytic expressions are available for $p(k)$ (see Ref.~\onlinecite{7} and references given there). 
In the limit of high temperature $\mu_D$ approaches the value $k_B T \ln{[n\lambda^D(T)/2]}$ 
where $\lambda(T)=\sqrt{2\pi\hbar^2/(mk_B T)}$ is the thermal de Broglie wavelength, 
and Eq.~(\ref{eq10}) yields the asymptotic result
\begin{equation}\label{eq11}
p(k)\rightarrow\frac{\Omega_D}{2^{D/2}\pi^D}\,k^{D-1}\lambda^{D}(T)
\,\exp{\left(-\frac{k^2\lambda^{2}(T)}{2\pi}\right)}\,.
\end{equation}
Numerical calculation is necessary at any finite temperature. In $D=2$ the specific form of the 
density of single-particle states yields the analytical result
\begin{equation}\label{eq12}
\mu_{2}=\varepsilon_{F}+k_B T\,\ln{[1-\exp{(-\varepsilon_{F}/k_B T)}]}\,.
\end{equation}

The dependence of $\mu_D$ on the reduced temperature $\Theta$ is shown in Fig. 1 at $r_s=5$ over a 
wide temperature range (recall that room temperature corresponds to $\Theta=0.025$). In Fig. 2 we 
show the momentum distribution $p(k)$ as a function of $k/k_F$ at $r_s=10$ in $D=3$ and at $r_s=5$ in $D=2$. 
The effects of temperature are evident from the Figure: (i) $p(k)$ acquires a high-momentum 
tail extending beyond the zero-temperature cut-off at $k=k_F$; and (ii) its peak moves to higher 
momenta from its zero-temperature position at $k=k_F/2$ in $D=3$ and $k \approx 0.42 k_F$ in $D=2$.

\section{Temperature dependence of pair distribution function}

We report in Figures 3 and 4 our numerical results for the temperature shifts of the pair 
distribution function in the $3D$ electron gas at $r_s=10$ and in the $2D$ gas at $r_s=5$. 
These shifts are qualitatively similar in both dimensionalities and can be said to be relatively small, 
if we bear in mind the magnitude of the unit of measure that we have adopted for the reduced temperature.

Evidently, these results reflect temperature-induced changes in the state of short-range 
order in the gas. Three ranges of values for the reduced distance $rk_F$ need to be separately 
examined: (i) the region of the main peak at $r_p$, lying near $r_pk_F \approx 3.8$ in $D=3$ and near $r_pk_F \approx 3$ in $D=2$, where thermal agitation is seen to reduce the magnitude of two-body correlations; (ii) 
the intermediate region around $r_pk_F/2$, where the conditioned probability of finding a second 
electron at distance $r$ from an electron at the origin is instead enhanced; and (iii) the region of 
close contact at distances approaching $r_pk_F \approx 0$, where again thermal agitation reduces the 
magnitude of $g(r)$.

The behaviour of the pair correlations in the first two regions above clearly reflects the 
increase in the mean kinetic energy with temperature, which is associated with the broadening of 
the distribution of relative momentum displayed in Figure 2. This enhances the tunnelling 
probability of a second electron into the potential barrier created by the electron at the origin and 
tends to smear out the first-neighbour shell at a given value of the coupling strength. Such a 
reduction in the state of short-range order leads us to expect that the phase boundary between 
the gas and the Wigner crystal will shift with increasing temperature to higher values of $r_s$ ({\it i.e.} 
to stronger effective Coulomb coupling).

The behaviour of $g(r)$ in the third region above was first noticed in the work of Macke and 
coworkers~\cite{4}, relating to the $3D$ gas in the weak-coupling regime. Their interpretation was that 
the width of the potential barrier surrounding the electron at the origin is also increasing with 
temperature, thus inhibiting deep penetration by a second electron. Our results in Figs. 3 and 4 
confirm this property of the electron gas as persisting in the intermediate-coupling regime and in 
lowered dimensionality.

\section{Summary and future perspectives}

In summary, we have solved the two-electron scattering problem in the electron gas at 
finite temperature within the self-consistent Hartree approximation, to display the changes in the 
state of short-range order that are brought about by thermal excitation.

The results achieved in previous work on pair correlations in the electron gas at zero 
temperature~\cite{7,11} immediately suggest what would be important improvements in the present 
approach, especially in regard to the $2D$ case and with a view to using the theory to construct 
reliable phase diagrams in a thermodynamic space spanned by density, temperature, and 
spontaneous spin polarization. Two directions seem worth exploring: (i) the inclusion of a spin 
dependence of the effective scattering potential, as developed in the work of Kukkonen and 
Overhauser~\cite{12}; and (ii) the derivation and study of differential equations obeyed directly by 
the pair distribution functions at finite temperature, rather than by their angular-momentum 
components.

As a final comment, we may recall that in the classical limit the density and temperature 
dependence of $g(r)$ are determined by three- and four-body distribution functions~\cite{13}. The 
expression for the temperature derivative of $g(r)$ can then be shown to relate to the heat capacity 
of the classical fluid {\it via} a fluctuation formula. The thermodynamics of a Fermi gas may also be 
worth studying from this viewpoint.

\begin{acknowledgments}
This work was partially supported by MIUR through the PRIN2003 program. MC 
acknowledges partial financial support by Fondecyt (Project 3010022) and a travel grant from 
INFM. MPT thanks Professor V.E. Kravtsov and the Condensed Matter Theory Group of the 
Abdus Salam ICTP for their hospitality during the final stages of this work.
\end{acknowledgments}

\newpage

\begin{figure}
\begin{center}
\includegraphics[scale=0.6]{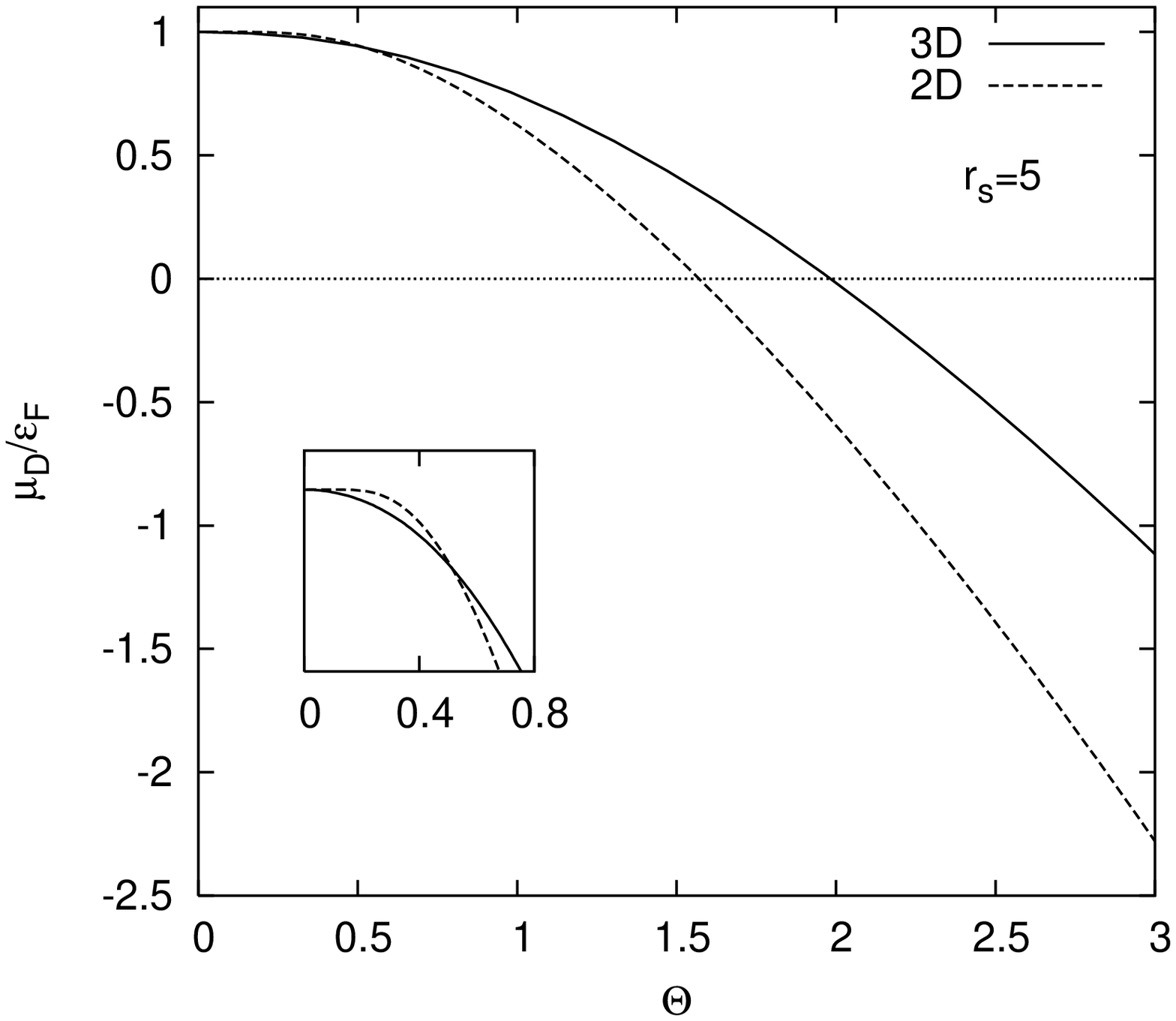}
\caption{Temperature dependence of the chemical potential of a Fermi gas at mean density 
corresponding to $r_s=5$ in $D=3$ and $D=2$: $\mu_D$ is plotted in units of $\varepsilon_F$ 
as a function of reduced temperature $\Theta$. 
The inset shows an enlargement of the results for $\Theta\leq 0.8$.}
\end{center}
\label{fig1}
\end{figure}

\begin{figure}
\begin{center}
\includegraphics[scale=0.45]{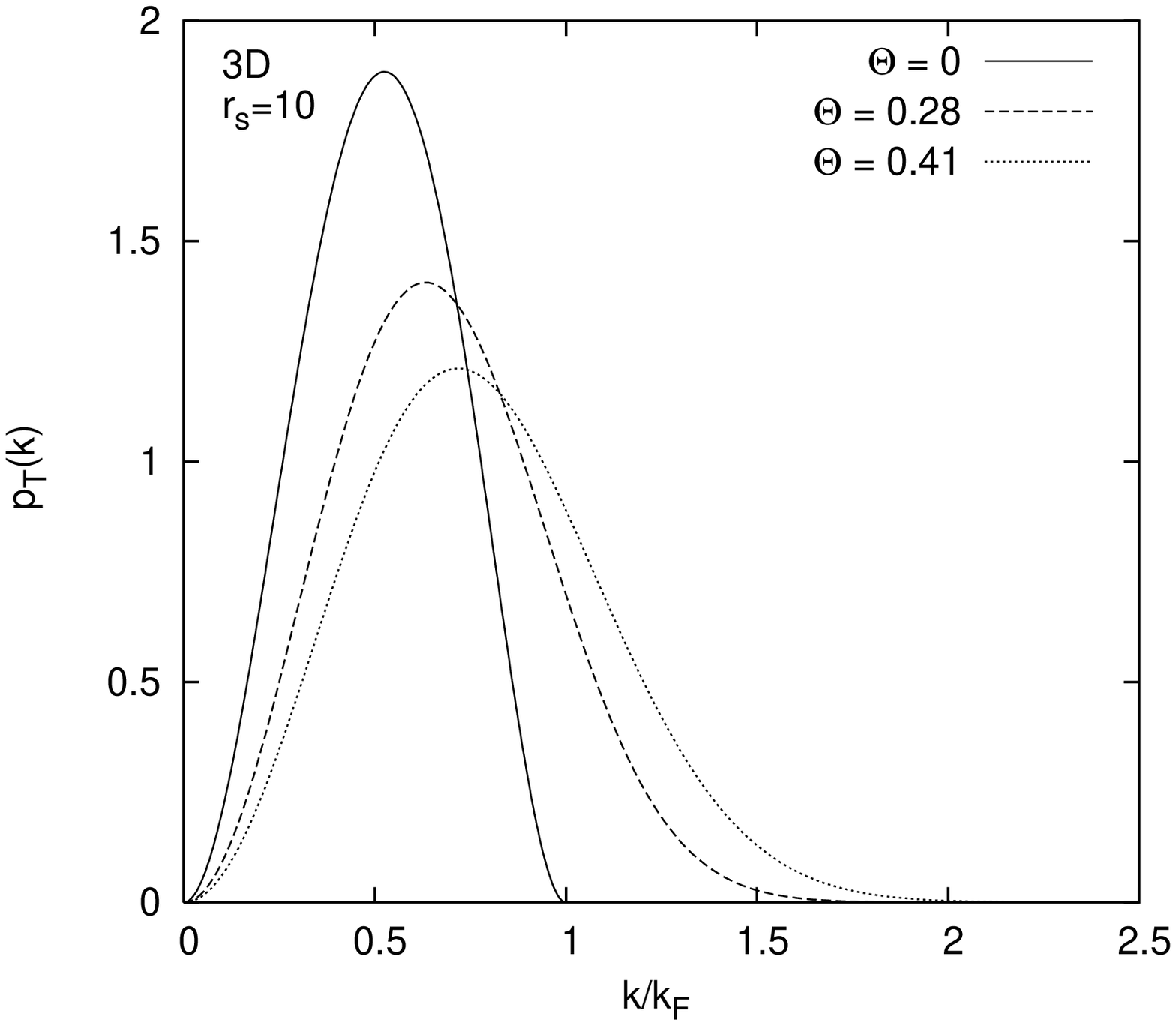}
\includegraphics[scale=0.45]{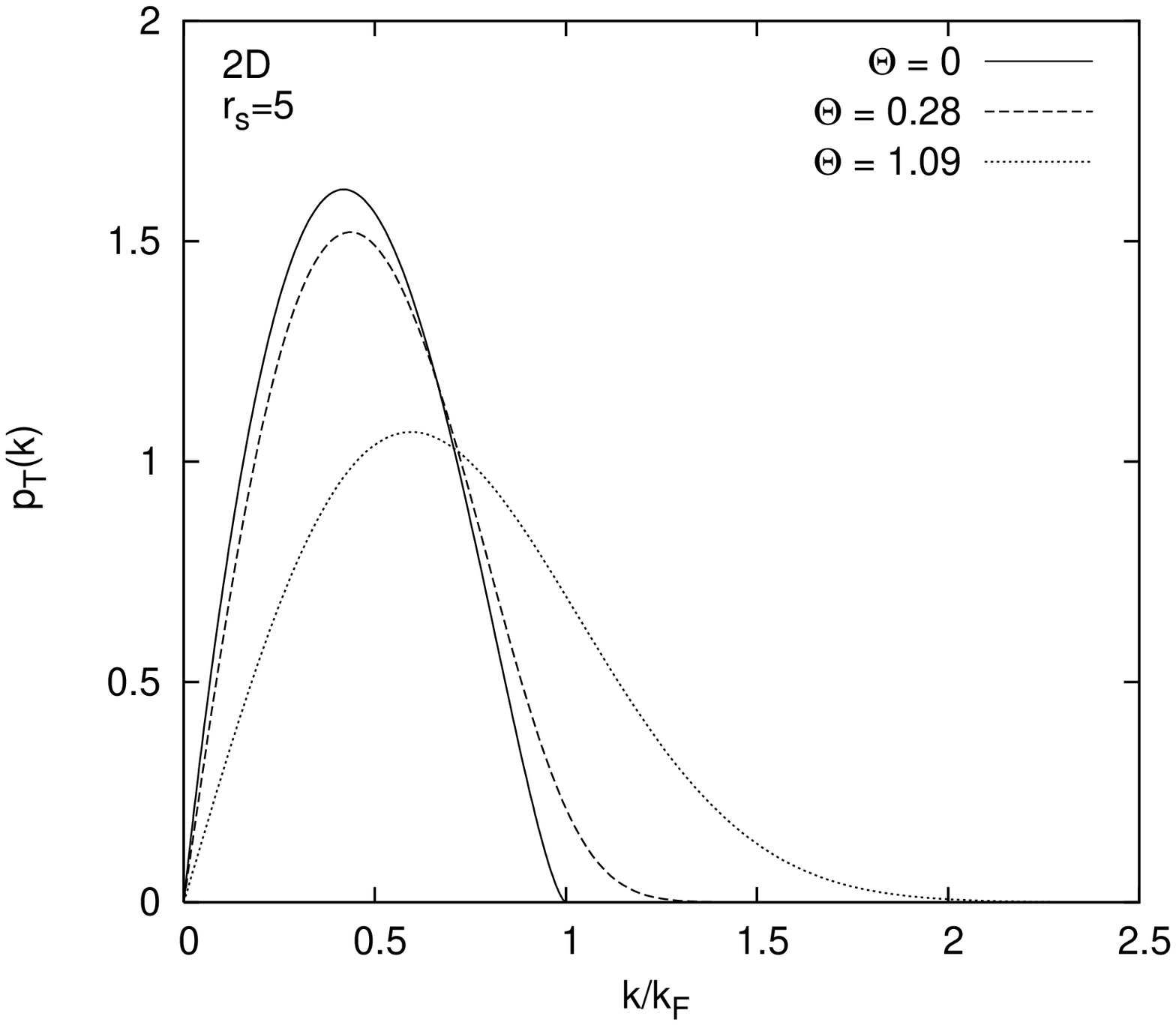}
\caption{Distribution $p(k)$ of relative momentum $k$ for an electron pair in a Fermi gas at the 
indicated values of the reduced temperature $\Theta$, as a function of $k/k_F$ for $r_s=10$ in $D=3$ (left 
panel) and for $r_s=5$ in $D=2$ (right panel).}
\end{center}
\label{fig2}
\end{figure}

\begin{figure}
\begin{center}
\includegraphics[scale=0.6]{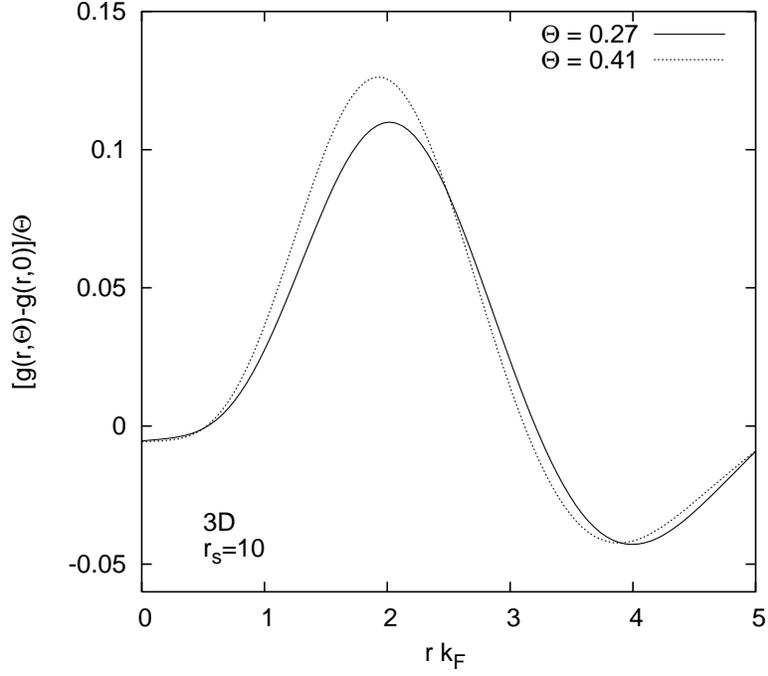}
\caption{Temperature dependence of the pair distribution function in the $3D$ electron gas at $r_s=10$: 
$[g(r,\Theta)-g(r,0)]/\Theta$ as a function of reduced distance $rk_F$ at two values of the reduced 
temperature $\Theta$.}
\end{center}
\label{fig3}
\end{figure}

\begin{figure}
\begin{center}
\includegraphics[scale=0.6]{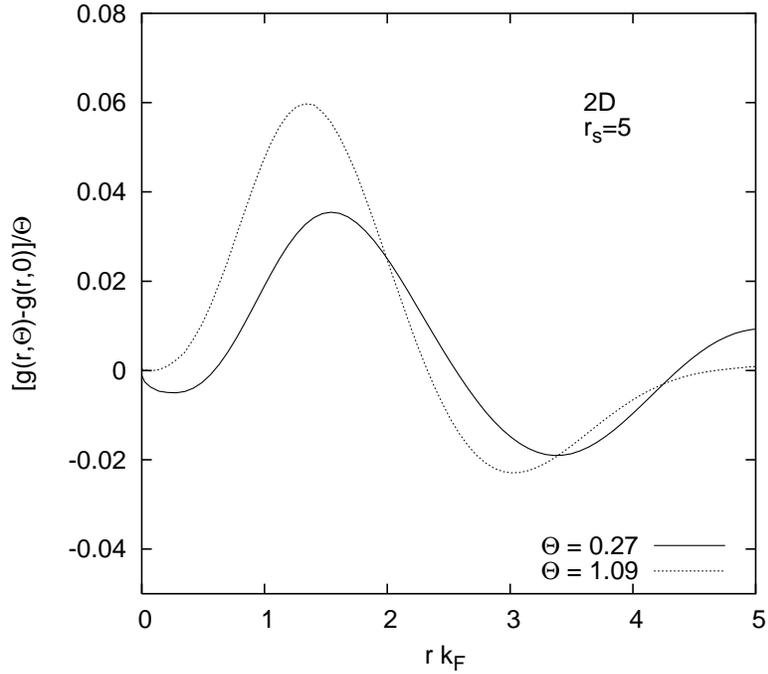}
\caption{Same as in Figure 3, for the $2D$ electron gas at $r_s=5$.}
\end{center}
\label{fig4}
\end{figure}

\end{document}